\documentstyle[preprint,aps]{revtex}

\begin{document}

\draft
\preprint{LBL-35980}

\title{The LPM effect in QCD and Radiative energy loss in a quark-gluon plasma}

\author{Xin-Nian Wang$^1$, Miklos Gyulassy$^2$ and Michael Pl\"umer$^{1,3}$}
\address{
$^1$Nuclear Science Division, Mailstop 70A-3307,
        Lawrence Berkeley Laboratory\\
University of California, Berkeley, California 94720.\\
$^2$Physics Department, Columbia University, New York, NY 10027\\
$^3$Physics Department, University of Marburg, D-35032 Marburg, Germany.
\thanks{Permanent address.} }
\date{August 18, 1994}
\maketitle
\begin{abstract}
The non-abelian analog of the Landau-Pomeranchuk-Migdal effect
is investigated in
perturbative QCD. Extending  our previous studies,
the suppression of induced soft bremsstrahlung due
to multiple scatterings of  quarks in the spinor
representation is considered. The
effective formation time of gluon radiation due to
the color interference is shown to depend on the color
representation of the emitting parton, and an improved
formula for the radiative energy loss is derived that interpolates
between the factorization and Bethe-Heitler limits.
\end{abstract}


\narrowtext

\section{INTRODUCTION}

Ultrarelativistic heavy ion collisions at future collider energies
of BNL Relativistic Heavy Ion Collider (RHIC) and CERN Large Hadron
Collider (LHC) are expected to reveal novel QCD
dynamics not accessible at present fixed target energies. When
the transverse momentum transfer involved in each nucleon-nucleon
collision is small, $p_T {\ \raisebox{2.75pt}{$<$}\hspace{-9.0pt}
\raisebox{-2.75pt}{$\sim$}\ }\Lambda_{\rm QCD}$, effective models
based on  meson-exchange and resonance formation
are sufficient to describe multiple
interaction between hadrons.
Those interactions lead to
collective behavior in low energy heavy ion collisions as  first
observed in Bevalac experiments\cite{BEVL}
and recently at AGS energies\cite{stachel}. When $p_T$ becomes
large enough to resolve individual partons inside a nucleon,
the dynamics is best described on the  parton level
via perturbative QCD. Though hard  parton interactions
occur at CERN-SPS energies ($E_{lab}<200$ AGeV),
they play a negligible role in the global
features of heavy ion collisions. However, at collider energies
($E_{cm}> 100$ AGeV) the importance of hard or semihard parton
scatterings  is  clearly seen in high-energy $pp$ and $p\bar{p}$
collisions \cite{WANGPP}. They are therefore also
expected to be dominant in heavy ion collisions at  RHIC and
LHC energies \cite{JBAM,KLL}.

Since hard parton scatterings occur  very early ($\sim 0.01$ fm/c)
and their rates are calculable via perturbative
QCD (pQCD), we proposed in \cite{MGMP,MGXW92}
that  high $p_T$ parton jets could serve as a unique
  probe of the quark-gluon plasma formed due to  copious
mini-jet production over  longer time scales ($>0.1$ fm/c).
The systematics of jet quenching  provides information
on  the stopping power, $dE/dz$, of high energy partons
in dense matter\cite{GPTW}.
The stopping power is in turn  controlled by the
color screening scale, $\mu$, in the medium.
Thus, jets provide information on that
interesting dynamical scale in deconfined matter.
At very high energy densities, that scale
is expected to be large
compared to confinement scale\cite{BMW}.
At temperatures, $T\gg T_c$, for example,
we expect $\mu\sim gT\gg \Lambda_{QCD}$.
In that case, most partonic interactions
have  high momentum transfers and perturbative
QCD methods may apply to the  calculation of  multiple collision
amplitudes.

In Ref.\cite{MGXW} a systematic study of QCD multiple collision
theory was initiated with the aim of deriving  the
non-abelian  analog of the
Landau-Pomeranchuk-Migdal (LPM) effect.
That effect, first derived in the case of QED\cite{LPM,SHUL} long ago,
involves the destructive interference between Bremsstrahlung radiation
amplitudes. It suppresses radiation
relative to the Bethe-Heitler formula in kinematic
regions where the radiation formation time is long compared
to the mean free path.
In QCD a similar effect is  also expected because it mainly
follows from general relativistic  uncertainty principle arguments.
However, we showed in \cite{MGXW} that specific non-abelian
effects influence the detailed interference pattern in the case
of QCD.

The LPM effect in QCD is especially important for  estimating  the
energy loss, $dE/dz$, of an energetic parton traversing a dense
QCD medium. We note that there exists considerable controversy
in the literature on the magnitude and energy dependence
of the energy loss  in QCD
\cite{RYSKI,GAVIN,SOREN,BRODS,RALSTON}.
In this paper we extend our derivation
in \cite{MGXW} to clarify further this effect
and improve  our previous estimates of $dE/dz$.

In the next section, we consider in detail the problem  of
induced  gluon radiation
from a spin 1/2 quark suffering two elastic scatterings.
This provides an insight into the applicability
of the effective potential model used in Ref.~\cite{MGXW}
to calculate multiple collision amplitudes.
We show why radiation from the target partons is negligible
although such amplitudes are absolutely necessary to
ensure gauge invariance. On the other hand, gauge
invariance constrains all gluon propagators to be
regulated by the same screening mass
in the  potential model. In section 3,
the  QCD radiation interference pattern is shown
to be expressible as a function of
 an effective formation time that depends on the color representation
of the jet parton.
In section 4, that formation time is used to calculate the average
radiative energy loss $dE/dz$ and derive
a simple formula that interpolates
between the  factorization and  Bethe-Heitler limits
as a function of a dimensionless ratio
depending on the  incident parton energy $E$, the
screening scale and the mean free path.
Finally, section 5 contains a summary and closing remarks.

\section{The Potential Model and Gauge Invariance}

To analyze multiple parton scatterings and the induced
gluon radiation, certain simplifications have to be made
for the interaction. Consider the scattering of a high
energy jet parton in a color neutral quark-gluon plasma.
If the average distance, $\Delta z=\lambda$,
between two successive scatterings is large compared
to the color screening length, $\lambda \gg \mu^{-1}$,
the effective average random color field produced by the
target partons can be modeled by a static Debye screened
potential:
\begin{equation}
V^a_{AA'}({\bf q})=A^a_{AA'}({\bf q})e^{-i{\bf q\cdot x}}
=gT^a_{AA'}\frac{e^{-i{\bf q\cdot x}}} {{\bf q}^2+\mu^2}, \label{eq:v1}
\end{equation}
where $\mu$ is the color screening mass, $T^a$ are the generators
of $SU(3)$ corresponding
to the representation of the target parton at ${\bf x}$.
The initial and final color indices, $A$, $A'$, of the target
parton are averaged and summed over when calculating the
ensemble averaged cross sections. This model potential was used in
Ref.~\cite{MGXW} to calculate cross sections of multiple scattering
and induced radiation. However, to obtain a gauge invariant amplitude
of gluon radiation in QCD, every diagram with the fixed number of
final gluon lines has to be taken into account, including gluon
radiation from the target parton line. Radiation
from target partons cannot be described by the
above static model potential.
The relative importance of different diagrams
to the net energy loss depends on the choice of the gauge.
In the light-cone gauge one expects that radiation
from target legs to be negligible as compared to
that from the high energy beam parton.

To estimate the importance of target radiation to the
energy loss of a fast parton and to see how the gauge invariance
constrains the potential model,
consider the simplest case of induced radiation from
quark-quark  scatterings. The Born amplitude for
$(p_i,k_i)\rightarrow (p_f,k_f)$ through one gluon
exchange is,
\begin{equation}
{\cal M}_{el}=ig^2T^a_{AA'}T^a_{BB'}\frac{\bar u(p_f)\gamma_{\mu}u(p_i)
        \bar u(k_f)\gamma^{\mu}u(k_i)}{(k_i-k_f)^2},
        \label{eq:el1}
\end{equation}
where $A$, $A'$, $B$, and $B'$ are the initial and final
color indices of the beam and target partons, respectively.
The corresponding elastic cross section is,
\begin{equation}
\frac{d\sigma_{el}}{dt}=C^{(1)}_{el}\frac{2\pi\alpha_s^2}{s^2}
                         \frac{s^2+u^2}{t^2},
\end{equation}
where $C^{(1)}_{el}=C_F/2N=2/9$ is the color factor for a single
elastic quark-quark scattering and $s$,$u$, and $t$ are the Mandelstam
variables.

For induced radiation, there are all together three groups of diagrams
as shown in Fig.~\ref{fig1}. If we rewrite the three amplitudes as
\begin{equation}
{\cal M}^{(i)}_{rad}\equiv \widehat{\cal M}^{(i)}_{\mu}\epsilon^{\mu},
\end{equation}
then gauge invariance implies that
\begin{equation}
\sum_{i=1}^3 \widehat{\cal M}^{(i)}_{\mu}k^{\mu}=0,
\end{equation}
where $\epsilon$ and $k$ are, respectively, the polarization and
momentum of the radiated gluon. In the soft radiation
limit, we can neglect all the terms proportional to $k_{\mu}$ in
$\widehat{\cal M}^{(i)}_{\mu}$ which does not contribute to
$\widehat{\cal M}^{(i)}_{\mu}k^{\mu}$($k^2=0$). The complete and
gauge invariant amplitude for induced gluon radiation is,
\begin{eqnarray}
{\cal M}_{rad} &=&ig^3\bar u(p_f)\gamma_{\mu}u(p_i)
 \bar u(k_f)\gamma^{\mu}u(k_f) \nonumber \\
 & &\left\{ \frac{1}{(k_i-k_f)^2}\left[\frac{\epsilon\cdot p_f}{k\cdot p_f}
 (T^bT^a)_{BB'}-\frac{\epsilon\cdot p_i}{k\cdot p_i}(T^aT^b)_{BB'} \right]
 T^a_{AA'}\right. \nonumber \\
 & &+\frac{1}{(p_i-p_f)^2}\left[\frac{\epsilon\cdot k_f}{k\cdot k_f}
 (T^bT^a)_{AA'}-\frac{\epsilon\cdot k_i}{k\cdot k_i}(T^aT^b)_{AA'} \right]
 T^a_{BB'} \nonumber \\
 & &+\left. \frac{1}{(p_i-p_f)^2(k_i-k_f)^2}[\epsilon\cdot(k_f-k_i)
 -\epsilon\cdot(p_f-p_i)]C^b_{AA',BB'}\right\}, \label{eq:rad1}
\end{eqnarray}
where $b$ is the color index of the radiated gluon, and
\begin{equation}
C^b_{AA',BB'}=[T^a,T^b]_{BB'}T^a_{AA'}
             =-T^a_{BB'}[T^a,T^b]_{AA'}
\end{equation}
are the color matrices associated with gluon radiation from the internal
gluon line (Fig.~\ref{fig1}c). The three terms in Eq.~(\ref{eq:rad1})
correspond to the radiation from the projectile (Fig.~\ref{fig1}a),
target parton (Fig.~\ref{fig1}b) and the internal gluon line
(Fig.~\ref{fig1}c).  As one can see, no  term alone is
separately invariant under a
gauge transformation $\epsilon^\mu\rightarrow
\epsilon^\mu +c\; k^\mu$. Only the total amplitude is gauge invariant.

In the potential model, one simply neglects the radiation from
the target lines and replace the gluon-exchange amplitude
associated with the target parton,
$gT^a_{AA'}\bar u(k_f)\gamma^{\mu}u(k_i)/(k_i-k_f)^2$,
by an effective potential $g^{\mu 0}A^a_{AA'}$. The potential
given by Eq.~(\ref{eq:v1}) is regularized by the
color screening mass squared, $\mu^2$. However
in Eq.~(\ref{eq:rad1}), if any of the gluon
propagators is regulated, gauge invariance is preserved only if all
propagators are regulated with the same mass squared, $\mu^2$.
In particular,  the two propagators in the
internal radiation diagram must both be regulated by the same
scale.

While gauge invariance is manifest only if all diagrams
are added, we now show that, in the $A^+=0$ gauge,
only the projectile diagrams contribute significantly
to the radiation of soft gluons in the dominant
kinematic range for the net energy loss.
We consider the case that a gluon with light-cone
momentum and polarization,
\begin{eqnarray}
k&=&[xP^+, k^2_{\perp}/xP^+, {\bf k}_{\perp}], \nonumber\\
\epsilon&=&[0,2\vec{\epsilon}_{\perp}\cdot{\bf k}_{\perp}/xP^+,
            \vec{\epsilon}_{\perp}],
\end{eqnarray}
is radiated off a high energy quark with initial momentum
\begin{equation}
p_i=[P^+,0,{\bf 0}_{\perp}].
\end{equation}
During the interaction, the beam quark exchanges a momentum
\begin{equation}
q=[q^+,q^-,{\bf q}_{\perp}]
\end{equation}
with a target quark which has a typical thermal momentum,
\begin{equation}
k_i=[M,M,{\bf 0}_{\perp}],
\end{equation}
with $M\sim T$ in the plasma rest frame.
 We focus on the limit defined by $x\ll 1$, however,
with the condition, $xP^+\gg M\gg q_{\perp}$.  By requiring both of
the final quarks to emerge on mass shell, one  finds that,
\begin{eqnarray}
q^+&\simeq&-\frac{q^2_{\perp}}{M}, \nonumber \\
q^-&\simeq&\frac{({\bf q}_{\perp}-{\bf k}_{\perp})^2}{(1-x)P^+}
            +\frac{k^2_{\perp}}{xP^+}.
\end{eqnarray}
The final momenta of the beam and target partons are, respectively,
\begin{eqnarray}
p_f&=&p_i+q-k\simeq [(1-x)P^+,\frac{({\bf q}_{\perp}-{\bf k}_{\perp})^2}
    {(1-x)P^+}, {\bf q}_{\perp}-{\bf k}_{\perp}],
    \nonumber \\
k_f&=&k_i-q\simeq [M+q^2_{\perp}/M,M,-{\bf q}_{\perp}].
\end{eqnarray}
With the above kinematics, one can obtain the momentum elements of the
radiation amplitudes:
\begin{equation}
\frac{\epsilon\cdot p_f}{k\cdot p_f}\simeq
\frac{\epsilon\cdot p_i}{k\cdot p_i}
=2\frac{\vec{\epsilon_{\perp}}\cdot {\bf k}_{\perp}}{k^2_{\perp}};
          \label{eq:ele1}
\end{equation}
\begin{equation}
\frac{\epsilon\cdot k_i}{k\cdot k_i}\simeq
2\frac{\vec{\epsilon_{\perp}}\cdot {\bf k}_{\perp}}{(xP^+)^2},\;\;
\frac{\epsilon\cdot k_f}{k\cdot k_f}\simeq
2\frac{\vec{\epsilon}_{\perp}\cdot {\bf q}_{\perp}}{xP^+M}.\label{eq:ele2}
\end{equation}
In the large $xP^+\gg k_{\perp}$ limit, we see that the magnitude
of the matrix elements involving target parton radiation
in Eq.~(\ref{eq:ele2}) are much smaller than those
involving projectile radiation in Eq.~(\ref{eq:ele1}).
As we will see below, the LPM effect limits the radiation
to $x< \lambda\mu^2/P^+$ and $k_\perp<\mu$. Therefore,
as long as $\lambda \mu\gg 1$, the contribution to the
energy loss in the main  kinematic range is dominated
by the projectile radiation in this gauge. However,
as we have demonstrated, the small target contributions
to the radiation amplitude are amplified
if $\epsilon^{\mu}$ is replaced by $k^{\mu}$. As a result of this
amplification, those amplitudes cannot be neglected
when considering gauge invariance even though they can be neglected
for calculating the energy loss.

Taking into account only the dominant contributions to the radiation
amplitude, we have the factorized amplitude,
\begin{eqnarray}
{\cal M}_{rad}&\equiv& \frac{{\cal M}_{el}}{T^a_{AA'}T^a_{BB'}}
                    i{\cal R}_1, \nonumber \\
{\cal R}_1&\simeq&2ig\vec{\epsilon}_{\perp}\cdot\left[
\frac{{\bf k}_{\perp}}{k^2_{\perp}}+\frac{{\bf q}_{\perp}-{\bf k}_{\perp}}
{({\bf q}_{\perp}-{\bf k}_{\perp})^2}\right]T^a_{AA'}[T^a,T^b]_{BB'},
\label{eq:rad2}
\end{eqnarray}
where ${\cal M}_{el}$ is the elastic amplitude as given in Eq.~(\ref{eq:el1}),
and ${\cal R}_1$ is defined as the radiation amplitude induced by a single
scattering. For later convenience, all the color matrix elements
are included in the definition of the radiation
amplitude ${\cal R}_1$.
With the above  approximations, we recover the differential
cross section for induced gluon bremsstrahlung by a single collision as
originally derived by Gunion and Bertsch~\cite{GUNION},
\begin{equation}
\frac{d\sigma}{dtdyd^2k_{\perp}}=\frac{d\sigma_{el}}{dt}
\frac{dn^{(1)}}{dyd^2k_{\perp}},
\end{equation}
where the spectrum for the radiated gluon is,
\begin{equation}
\frac{dn^{(1)}}{dyd^2k_{\perp}}\equiv\frac{1}{2(2\pi)^3C^{(1)}_{el}}
\overline{\left|{\cal R}_1\right|^2}
=\frac{C_A\alpha_s}{\pi^2}\frac{q^2_{\perp}}
{k^2_{\perp}({\bf q}_{\perp}-{\bf k}_{\perp})^2}. \label{eq:spec1}
\end{equation}
In the square modulus of the radiation amplitude, an average and a sum
over initial and final color indices and polarization are understood.
We see that the spectrum has a uniform distribution in central rapidity
region (small $x$) which is a well known feature of QCD soft
radiation \cite{DKMT}.
This feature is consistent with the hadron distributions predicted by
Lund string models and the ``string effects'' in $e^+e^-$ three jets
events \cite{LUND,NIED}, all being the results of interference effects of
pQCD radiation. To demonstrate this a little in detail, let us
consider only the radiation amplitude from the beam quark in
Eq.~(\ref{eq:rad1}),
\begin{equation}
{\cal R}=\left[\frac{\epsilon\cdot p_f}{k\cdot p_f}(T^bT^a)_{BB'}
-\frac{\epsilon\cdot p_i}{k\cdot p_i}(T^aT^b)_{BB'} \right]T^a_{AA'}.
\end{equation}
The corresponding gluon spectrum is (beside a factor $1/2(2\pi)^3$),
\begin{equation}
\frac{1}{C^{(1)}_{el}}\overline{\left|{\cal R}\right|^2}
=C_F\left[\frac{\epsilon\cdot p_i}{k\cdot p_i}
-\frac{\epsilon\cdot p_f}{k\cdot p_f}\right]^2
+\frac{C_A}{2}2\frac{\epsilon\cdot p_i}{k\cdot p_i}
\frac{\epsilon\cdot p_f}{k\cdot p_f},
\end{equation}
where $C_F=(N^2-1)/2N$ and $C_A=N$ are the second order
Casimir for quarks in the fundamental and for gluons in the adjoint
representation, respectively. Note that the first term is identical
to  gluon radiation induced by an abelian gauge interaction
like a photon exchange
and does not contribute to the gluon spectrum in central rapidity region
due to the destructive interference of the initial and final state
radiations.
The second term arises from the non-abelian
interactions with the target partons
and is the main contribution to the central region.

Another nonabelian feature in the induced gluon radiation amplitude,
Eq.~(\ref{eq:rad2}), is the singularity at ${\bf k}_{\perp}={\bf q}_{\perp}$
due to induced radiation along the direction of the exchanged gluon.
For $k_{\perp}\ll q_{\perp}$, we note that the induced radiation
from a three gluon vertex can be neglected as compared to the leading
contribution $1/k^2_{\perp}$. However, at large $k_{\perp}\gg q_{\perp}$,
this three gluon amplitude is important to change the gluon spectrum
to a $1/k^4_{\perp}$ behavior, leading to a finite average transverse
momentum. Therefore, $q_{\perp}$ may serve as a cut-off for $k_{\perp}$
when one neglects the amplitude with the three gluon vertices as we
will do when we consider induced radiation by multiple scatterings
in the next section. If one wishes to include the three gluon amplitude,
then the singularity at ${\bf k}_{\perp}={\bf q}_{\perp}$ has to
be regularized. As we have discussed above, the regularization
scheme has to be the same as for the model potential or the gluon
propagator in $d\sigma_{el}/dt$ as required by gauge invariance.
In our case, a color screening mass $\mu$ will be used.

\section{EFFECTIVE FORMATION TIME}

The radiation amplitude induced by multiple scatterings has been
discussed in Ref.~\cite{MGXW}. We  discuss here the
special case of  double scatterings to gain further
insight into the problem. Consider two static
potentials separated by a distance $L$, which is assumed
to be much larger than
the interaction length, $1/\mu$.
In the abelian case, the radiation amplitude associated with double
scatterings is (see Appendix),
\begin{equation}
{\cal R}_2^{\rm QED}=ig\left[\left(\frac{\epsilon\cdot p_i}{k\cdot p_i}
-\frac{\epsilon\cdot p}{k\cdot p}\right)e^{ik\cdot x_1}+\left(
\frac{\epsilon\cdot p}{k\cdot p}-\frac{\epsilon\cdot p_f}{k\cdot p_f}
\right)e^{ik\cdot x_2}\right],\label{eq:radQED}
\end{equation}
where $p=(p_f^0,p_z,{\bf p}_{\perp})$ is the four-momentum of
the intermediate parton line which is put on mass shell by the pole
in one of the parton propagators, $x_1=(0,{\bf x}_1)$, and
$x_2=(t_2,{\bf x}_2)$ are the four-coordinates of the two potentials
with $t_2=(z_2-z_1)/v_z=Lp^0/p_z$. This formula has been recently used
to discuss the interference effects on photon and dilepton production
in a quark gluon plasma \cite{CLEY,KNLL}. We notice that the amplitude has
two distinct contributions from each scattering. Especially,
the diagram (Fig.~\ref{fig2}b) with a gluon radiated from the intermediate
line between the two scatterings contributes both as the final
state radiation for the first scattering and the initial state
radiation for the second scattering.  The relative phase factor
$$k\cdot(x_2-x_1)=\omega(1/v_z-\cos \theta)L\equiv L/\tau(k)$$
determines the interference between radiations from the
two scatterings and is simply the ratio of the path length
to the formation time defined as
\begin{equation}
\tau(k)=\frac{1}{\omega(1/v_z-\cos \theta)}\simeq
\frac{2\omega}{k^2_{\perp}}.
\label{abtau}
\end{equation}
The Bethe-Heitler limit is reached when $L\gg \tau(k)$. In this
limit, the intensity of induced radiation is  additive in the
number of scatterings. However, when $L\ll\tau(k)$, the final
state radiation amplitude from the first scattering is
mostly cancelled by
the initial state radiation amplitude from the second scattering. The
radiation pattern then looks as if the parton has only suffered a single
scattering from $p_i$ to $p_f$.
This destructive interference is
 often referred to as the Landau-Pomeranchuk-Migdal
(LPM) effect. The corresponding limit is usually called the
factorization limit.

The radiation amplitude in QCD is similar to Eq.~(\ref{eq:radQED}),
except that one has to include different color factors for each
diagram in Fig.~\ref{fig2}. In the high energy limit,
$\epsilon\cdot p_i/k\cdot p_i\simeq \epsilon\cdot p/k\cdot p
\simeq \epsilon\cdot p_f/k\cdot p_f\simeq 2\vec{\epsilon}_{\perp}
\cdot{\bf k}_{\perp}/k^2_{\perp}$. The momentum dependence of each
contribution can be factorized out and the radiation amplitude for
diagrams in Fig.~\ref{fig2} is,
\begin{equation}
{\cal R}_2=i2g\frac{\vec{\epsilon}_{\perp}\cdot{\bf k}_{\perp}}
{k^2_{\perp}}\left\{\left(T^{a_2}[T^{a_1},T^b]\right)_{BB'}
e^{ik\cdot x_1}+\left([T^{a_2},T^b]T^{a_1}\right)_{BB'}
e^{ik\cdot x_2}\right\}T^{a_1}_{A_1A_1'}T^{a_2}_{A_2A_2'}, \label{eq:radQCD}
\end{equation}
where we have included two color matrices from the potentials, and
$b$ again represents the color index of the radiated gluon.
The radiation amplitude from diagrams
with three gluon  vertices has  the same phase and color
structures as in Eq.~(\ref{eq:radQCD}), but the momentum
dependence cannot be factorized, since each term depends on the
transverse momentum transfer which differs from one scattering
to another. However, since we are  interested in the soft radiation
limit, $k_\perp < q_\perp\sim \mu$,
the  contributions from internal gluon
line emissions can be neglected as shown in \cite{MGXW}.
 As we discussed above, however, those amplitudes serve to
provide an effective cutoff $\langle q_{\perp}\rangle\sim \mu$ for
$k_{\perp}$.

The extrapolation of Eq.~(\ref{eq:radQCD}) to the general case of
$m$ number of scatterings is straightforward
with the result
\begin{equation}
{\cal R}_m=i2g\frac{\vec{\epsilon}\cdot{\bf k}_{\perp}}{k^2_{\perp}}
T^{a_1}_{A_1A_1'}\cdots T^{a_m}_{A_mA_m'}\sum_{i=1}^m
\left(T^{a_m}\cdots[T^{a_i},T^b]\cdots T^{a_1}\right)_{BB'}e^{ik\cdot x_i},
\label{eq:radm}
\end{equation}
The above amplitude
contains $m$ terms each having a common momentum dependence in the high
energy limit, but with different color and phase factors. The above
expression is also valid for a gluon beam jet, with the corresponding
color matrices replaced by those of an adjoint representation. In
Eq.~(\ref{eq:radm}), we also assumed that all the potentials have a color
structure of a fundmental representation. One can generalize to the case
in which each individual potential could have any arbitrary color
representation. However, our following results on the gluon spectrum
and interference pattern will remain the same. With this in mind, we
have the spectrum of soft bremsstrahlung associated with multiple
scatterings in a color neutral ensemble, similar to Eq.~(\ref{eq:spec1}),
\begin{equation}
\frac{dn^{(m)}}{dyd^2k_{\perp}}=\frac{1}{2(2\pi)^2C^{(m)}_{el}}
\overline{|{\cal R}_m|^2}\equiv C_m(k)\frac{dn^{(1)}}{dyd^2k_{\perp}},
\label{eq:specm}
\end{equation}
where $C^{(m)}_{el}=(C_F/2N)^m$ is the color factor for
the elastic scattering cross section without radiation. $C_m(k)$,
defined as the ``radiation formation factor'' to characterize the
interference pattern due to multiple scatterings, can be expressed as
\begin{equation}
C_m(k)=\frac{1}{C^m_FC_AN}\sum_{i=1}^m\left[C_{ii}
+2Re\sum_{j=1}^{i-1}C_{ij}e^{ik\cdot(x_i-x_j)}\right],
\end{equation}
where the color coefficients, as computed  in Ref. \cite{MGXW}, are
\begin{eqnarray}
C_{ii}&=&C^m_FC_AN; \nonumber \\
C_{ij}&=&-\frac{C_A}{2}\frac{C_A}{2C_F}(1-\frac{C_A}{2C_F})^{i-j-1}
C^m_FC_AN,\;\; {\rm for }\; j<i.
\end{eqnarray}
For a gluon beam jet, one can simply change the dimension to that of an
$SU(N)$ adjoint representation and replaces $C_2=C_F$ by the corresponding
second order Casimir $C_2=C_A$. We then obtain a general form for the
radiation formation factor for a high energy parton jet,
\begin{equation}
C_m(k)=m-r_2Re\sum_{i=1}^m\sum_{j=1}^{i-1}(1-r_2)^{i-j-1}
e^{ik\cdot(x_i-x_j)}, \label{eq:radf}
\end{equation}
where
\begin{equation}
r_2=\frac{C_A}{2C_2}=\left\{ \begin{array}{ll}
N^2/(N^2-1) & \mbox{for quarks with}\; C_2=C_F\\
1/2         & \mbox{for gluons with}\; C_2=C_A \end{array} \right. .
\end{equation}

Similar to the special case of double scatterings, there are a few
interesting limits for the above general form of radiation formation
factor and the induced gluon spectrum. When $m=1$, $C_1(k)=1$. We
recover the gluon spectrum induced by a single scattering
in Eq.~(\ref{eq:spec1}) in the small $k_{\perp}$ limit. For multiple
scatterings in the Bethe-Heitler limit
when $L_{ij}=|z_i-z_j| \gg \tau (k)$
for all $i>j$, the phase factors average to zero and the intensity
of the radiation is additive in the number of scatterings, i.e.,
$C_m(k)\approx m$.
In the factorization limit, one has $L_{ij}\ll \tau(k)$ for all
$i>j$. In this case, the phase factors can be set to unity and
the summations in Eq.~(\ref{eq:radf}) can be carried out to give
\begin{equation}
C_m(k)\approx \frac{1}{r_2}[1-(1-r_2)^m]
=\left\{ \begin{array}{ll} 8/9[1-(-1/8)^m] \; & \mbox{for quarks} \\
2(1-1/2^m)& \mbox{for gluons} \end{array} \right. .
\end{equation}
In contrast to the Bethe-Heitler limit, the factorization limit is
independent of the number of collisions as $m\rightarrow\infty$,
and the radiation formation factor approaches $1/r_2=2C_2/C_A$.
It is interesting to note that the destructive interference for
quarks in the fundamental representation is so effective that
the radiation spectrum induced by many scatterings is even slightly less,
$1/r_2=8/9$, than by a single scattering. For gluon jets, however, the
interference is not as complete as for quarks. The induced radiation
approaches 2 times that from a single scattering. Using these values
of $C_m(k)=1/r_2=2C_2/C_A$ in Eqs.~(\ref{eq:specm}) and (\ref{eq:spec1}),
the radiation intensity induced by multiple scatterings is proportional
to $2C_2$ as compared to $C_A$ in the single scattering case.
The gluon intensity radiated by a gluon jet is therefore $9/4$ higher
than that by a quark due to the interference in multiple scatterings.
This dependence of LPM effect in QCD on the color representation of the
beam parton is a unique non-abelian effect. As we will discuss in the
following, the effective formation time of the radiation in a QCD
medium should also take this non-abelian effect into account.

To see analytically how $C_m(k)$ interpolates between the Bethe-Heitler
and factorization limits, let us average over the interaction points
${\bf x}_i$ according to a linear kinetic theory. We take an eikonal
approximation \cite{MGXW} for the multiple scatterings so that the
transverse phase factors can be neglected in the soft radiation limit,
$k_{\perp}\ll q_{\perp}\sim \mu$. In a linear kinetic theory, the
longitudinal separation between successive scatterings, $L_i=z_{i+1}-z_i$,
has a distribution,
\begin{equation}
\frac{dP}{dL_i}=\frac{1}{\lambda}e^{-L_i/\lambda},
\end{equation}
which is controlled by the mean free path, $\lambda$, of the scatterings.
The averaging of the phase factors,
\begin{equation}
  \left\langle e^{ik\cdot (x_i-x_j)}\right\rangle\approx
\left[\frac{1}{1-i\lambda/\tau(k)}\right]^{i-j},
\end{equation}
enables us to complete the summation in Eq.~(\ref{eq:radf}).
Neglecting terms proportional to $(1-r_2)^m$ for relatively
large $m$, we have,
\begin{equation}
C_m(k)\approx m\frac{\chi^2(k)}{1+\chi^2(k)}+
\frac{1-(1-2r_2)\chi^2(k)}{r_2[1+\chi^2(k)]^2}.
\label{cm}\end{equation}
The non-abelian LPM effect in QCD is therefore controlled
by the dimensionless ratio of the mean free path to an effective
formation time,
\begin{equation}
\chi(k)=\lambda/\tau_{\rm QCD}(k),
\end{equation}
where  the effective formation time in QCD
depends on the color representation of the jet parton
and is related to the
usual abelian formation time in Eq.(\ref{abtau}) as,
\begin{equation}
\tau_{\rm QCD}(k)=r_2\tau(k)=\frac{C_A}{2C_2}\frac{2\cosh y}{k_{\perp}},
                            \label{eq:ftime}.
\end{equation}
This formula for the radiation formation factor
is illustrated  in Fig. \ref{fig3}
as a function of $\tau(k)/\lambda$ for the case
of five collisions $(m=5)$ and shows
how $C_m(k)$ interpolates between the Bethe-Heitler limit
for small value of $\tau(k)/\lambda$ and the factorization
limit for large value of $\tau(k)/\lambda$.
For radiation with  average transverse momentum $k_\perp\sim \mu$
the additive Bethe-Heitler region is limited to rapidities
$y < \log(r_2 \lambda \mu)$.
In general, the radiation formation factor
is proportional to $m$ in the limit of large $m$. We can therefore regard
the radiation as being additive to the number of scatterings
with the radiation from each scattering
simply  suppressed by the factor $\chi^2/(1+\chi^2)$ due
to nonabelian LPM effect.
Since this effective formation time is a result of the
unique color interference effect in QCD, we should
use it in the following to estimate the
radiative energy loss by a high energy parton traversing a
color neutral quark-gluon plasma.

\section{RADIATIVE ENERGY LOSS}

We  now apply the effective formation time to derive a simple
approximate formula for the induced radiative
energy loss extending our previous result in \cite{MGXW}.
As shown in the previous section the radiation spectrum is
given by Eq.~(\ref{eq:spec1}) multiplied  by the
radiation formation factor Eq.~(\ref{cm}).
That factor simply restricts the additive kinematic region
to $\tau_{\rm QCD}(k)/\lambda < 1$ and the
the incremental change in that factor for each successive
collision can be approximated by a pocket formula
$dC_m(k)/dm\approx \theta(\lambda-\tau_{QCD}(k))$ (see Fig.3).
The  additive radiative energy loss for each  collision
beyond the first one is then
\begin{equation}
\Delta E_{rad}\approx\int d^2k_{\perp}dy\frac{dn_g}{d^2k_{\perp}dy}
k_{\perp}\cosh y\;\theta(\lambda-\tau_{\rm QCD}(k))
\; \theta(E-k_{\perp}\cosh y), \label{eq:de1}
\end{equation}
where $\tau_{\rm QCD}(k)$ is given by Eq.~(\ref{eq:ftime}), the second
$\theta$-function is for energy conservation, and the regularized gluon
density distribution induced by a single scattering is,
\begin{equation}
\frac{dn_g}{d^2k_{\perp}dy}=\frac{C_A\alpha_s}{\pi^2}
\frac{q^2_{\perp}}{k_{\perp}^2
[({\bf q}_{\perp}-{\bf k}_{\perp})^2+\mu^2]}. \label{eq:reg}
\end{equation}
As  discussed earlier, gauge invariance requires that the
singularity at ${\bf k}_{\perp}={\bf q}_{\perp}$ in Eq.~(\ref{eq:spec1})
must  be regularized by the same color screening mass $\mu$
as in the elastic cross section in the potential model.
Since the transverse momentum transfer $q_{\perp}$ is the result of
 elastic scatterings, we have to  average any function
$f({\bf q}_{\perp})$ of ${\bf q}_{\perp}$ by the elastic cross section,
\begin{equation}
\langle f({\bf q}_{\perp})\rangle=\frac{1}{\sigma_i}
\int_{\mu^2}^{s/4}dq^2_{\perp}
\frac{d\sigma_i}{dq^2_{\perp}}f({\bf q}_{\perp}),
\end{equation}
where $s\approx 6ET$ is the average $c.m.$ energy squared for the
scattering of a jet parton with energy $E$ off the thermal partons
at temperature $T$. For the dominant small angle scattering,
the elastic cross sections are,
\begin{equation}
\frac{d\sigma_i}{dq^2_{\perp}}\cong C_i
\frac{2\pi \alpha^2_s}{q^4_{\perp}}, \label{eq:els}
\end{equation}
where $C_i=9/4,\;1,\;4/9$ respectively for $gg$, $gq$ and $qq$ scatterings.
This average can be approximated by
replacing $q^2_{\perp}$ in the numerator of Eq.~(\ref{eq:reg})
with its average value,
\begin{equation}
  \langle q^2_{\perp}\rangle=\mu^2\ln\frac{3ET}{2\mu^2}. \label{eq:avq}
\end{equation}
In the denominator, we simply replace ${\bf q_{\perp}^2}$
by $\mu^2$ after the angular integration.
The remaining  integration in Eq.~(\ref{eq:de1}) over the restricted
phase space approximately leads to the simple analytic formula
\begin{equation}
\Delta E_{rad}\approx \frac{C_A\alpha_s}{\pi}
\langle q^2_{\perp}\rangle\left(\frac{\lambda}{2r_2}I_1
+\frac{E}{2\mu^2}I_2\right), \label{eq:de2}
\end{equation}
\begin{eqnarray}
I_1&=&\ln\left[\frac{r_2E}{\mu^2\lambda}+
\sqrt{1+\left(\frac{r_2E}{\mu^2\lambda}\right)^2}\,\right]-
\ln\left[2\left(\frac{r_2}{\mu\lambda}\right)^2+
\sqrt{1+4\left(\frac{r_2}{\mu\lambda}\right)^4}\,\right], \label{eq:I1} \\
I_2&=&\ln\left[\frac{\mu^2\lambda}{r_2E}+
\sqrt{1+\left(\frac{\mu^2\lambda}{r_2E}\right)^2}\,\right]-\ln\left[
\frac{2\mu^2}{E^2}+\sqrt{1+\left(\frac{2\mu^2}{E^2}\right)^2}\,\right]
.\label{eq:I2}
\end{eqnarray}
In small $k_{\perp}$ regime, the phase space is mainly restricted by
a small effective formation time, $\tau_{\rm QCD}<\lambda$, which
gives the first term proportional to $\lambda$. For large $k_{\perp}$,
the radiation becomes additive in a restricted
phase space constrained by energy conservation.
That region  contributes to the
second term which appears to be proportional to the incident energy $E$.
However, in the high energy limit the function $I_2\propto 1/E$ and hence
the radiated energy loss grows only as $\log^2 E$.

The above derivation assumed that the mean free path
is much larger than the interaction
range specified by $1/\mu$. As we shall discuss below,
this is satisfied in a quark gluon plasma
 at least in the weak coupling limit. Therefore, we can
neglect the second term in $I_1$. For a high energy jet parton,
$E\gg\mu$, we can also neglect the second term in $I_2$. The resulting
radiative energy loss reduces in that case to the simple form,
\begin{equation}
\frac{dE_{rad}}{dz}=\frac{\Delta E_{rad}}{\lambda}\approx
\frac{C_2\alpha_s}{\pi}\langle q^2_{\perp}\rangle
\left[\ln\left(\xi+\sqrt{1+\xi^2}\right)+\xi
\ln\left(\frac{1}{\xi}+\sqrt{1+\frac{1}{\xi^2}}\right)\right],
\label{eq:dedz1}
\end{equation}
which depends on a dimensionless variable,
\begin{equation}
\xi=\frac{r_2E}{\mu^2\lambda}.
\end{equation}
Since we have used gluon spectrum from a single scattering in
Eq.~(\ref{eq:spec1}) which is valid for all values of $k_{\perp}$,
the full integration over $k_{\perp}$ results in the logarithmic
energy dependence of $dE_{rad}/dz$. This logarithmic dependence
was absent in our derivation in Ref.\cite{MGXW}
since there only the contribution from
the very soft $k_{\perp}<\mu$ region was considered.

We  see that the radiative energy loss $dE_{rad}/dz$ thus obtained
interpolates between the factorization and Bethe-Heitler
limits as a function of the dimensionless ratio $\xi$. In the factorization
limit, we fix $\mu\lambda\gg 1$ and let $E\rightarrow\infty$,
so that $\xi\gg 1$. In this case, we can neglect the second term
in Eq.~(\ref{eq:dedz1}) and have,
\begin{equation}
  \frac{dE_{rad}}{dz}\approx \frac{C_2\alpha_s}{\pi}
 \langle q^2_{\perp}\rangle\ln\left(\frac{2r_2E}{\mu^2\lambda}\right);\;\;
 \xi\gg 1. \label{eq:dedzf}
\end{equation}
Thus, the radiative energy loss in the factorization limit has only
a logarithmic energy dependence (in addition to the energy dependence
of $\langle q_{\perp}^2\rangle$). Due to the non-abelian nature of the
color interference, the resultant energy loss for a gluon ($C_2=C_A)$
is 9/4 times larger than that for a quark ($C_2=C_F$). In the other
extreme limit, we fix $E$ and let $\mu\lambda\rightarrow\infty$, so that
$\xi\ll 1$. In this case, the mean free path exceeds the effective
formation time. The radiation from each scattering adds up. We
then recover the linear dependence of the energy loss $dE_{rad}/dz$
on the incident energy $E$ (modulo logarithms),
\begin{equation}
\frac{dE_{rad}}{dz}\approx  \frac{C_A\alpha_s}{2\pi\lambda}
\frac{\langle q^2_{\perp}\rangle}{\mu^2}
E\ln\left(\frac{2\mu^2\lambda}{r_2E}\right);\;\; \xi\ll 1, \label{eq:dedzb}
\end{equation}
as in the Bethe-Heitler formula. In both cases, the radiative energy
loss is proportional to the average of the transverse momentum transfer,
$\langle q_{\perp}^2\rangle$, which is controlled by
the color screening mass as in Eq.~(\ref{eq:avq}).

To see  more clearly how  the factorization limit
is approached, we now  estimate $\xi$ for a parton propagating
inside a high temperature quark-gluon plasma. From
Eq.~(\ref{eq:els}) and the perturbative QCD expressions
for the quark and gluon densities at temperature, $T$,
the mean free path for  3 quark flavors is
\begin{equation}
\lambda^{-1}_q=\sigma_{qq}\rho_q+\sigma_{qg}\rho_g\approx
\frac{2\pi\alpha^2_s}{\mu^2}4\times 7\zeta(3)
\frac{T^3}{\pi^2},\label{eq:lam1}
\end{equation}
\begin{equation}
\lambda^{-1}_g=\sigma_{qg}\rho_q+\sigma_{gg}\rho_g\approx
\frac{2\pi\alpha^2_s}{\mu^2}9\times 7\zeta(3)
\frac{T^3}{\pi^2}, \label{eq:lam2}
\end{equation}
where $\zeta(3)\approx 1.2$.
We emphasize that the above mean free path corresponds
approximately  to the color relaxation mean free
path, $\lambda_c$, and not the momentum relaxation
mean free path, $\lambda_p$. As shown in Ref.\cite{selik},
$\lambda_c\sim \alpha_s \lambda_p$ is generally
the shorter of the two in QCD.
The reason why the color relaxation mean free path
controls the radiation pattern is that the color current
responsible for emitting the gluons is coherent only
over a distance scale $\lambda_c$. It take a much longer
path length to stop a parton. However, unlike in QED, this
longer momentum relaxation mean free path is irrelevant
for nonabelian radiation.

Using Eqs. (\ref{eq:lam1}) and (\ref{eq:lam2})
 and the perturbative color electric screening mass,
$\mu^2=4\pi\alpha_sT^2$, we see that  $\xi$ appearing in the logarithms has a
common energy and temperature dependence for both quarks and gluons,
\begin{equation}
\xi=\frac{r_2E}{\lambda\mu^2}=\frac{63\zeta(3)}{16\pi^3}\frac{E}{T}
\approx\frac{9}{2\pi^3}\frac{E}{T},
\;\;\mbox{for both $q$ and $g$}. \label{eq:xi}
\end{equation}

With the above expression for $\xi$, we plot the radiative energy
loss in Fig.~\ref{fig4} as a function of the beam energy inside a plasma
at temperature $T=300$ MeV with $\alpha_s=0.3$.
The solid line is the full expression in Eq.~(\ref{eq:dedz1}) while
the dashed line is the factorization limit corresponding to the first
term in Eq.~(\ref{eq:dedz1}). We see that Eq. (\ref{eq:dedzf})
approximates Eq. (\ref{eq:dedz1}) quite well in this parameter range.
The energy dependence of the radiative energy loss
is due to the double logarithmic function in the formula one of which
comes from the energy dependence of the average transverse momentum
$\langle q^2_{\perp}\rangle$ in Eq.~(\ref{eq:avq}).

The energy loss of a quark in dense matter due to elastic scattering
was first estimated by Bjorken \cite{BJ} and later was studied in detail
\cite{THOMA} in terms of finite temperature QCD.  For our purpose, a
simple estimate taking into account both the thermal average and
color screening will suffice. In terms of elastic cross sections
and the density distributions for quarks and gluons in a plasma,
we have,
\begin{equation}
\frac{dE_{el}}{dz}=\int_{\mu^2}^{s/4}dq^2_{\perp}
\frac{d\sigma_i}{dq^2_{\perp}}\rho_i\nu=\langle q^2_{\perp}\rangle
\sigma_i\langle \frac{\rho_i}{2\omega}\rangle,
\end{equation}
where $\nu\approx q^2_{\perp}/2\omega$ is the energy transfer of the
jet parton to a thermal parton with energy $\omega$ during an elastic
scattering, $\langle q^2_{\perp}\rangle$ is the average transverse
momentum transfer given by Eq.~(\ref{eq:avq}). Similar to
Eqs.~(\ref{eq:lam1}) and (\ref{eq:lam2}), we have,
\begin{equation}
\sigma_{qq}\langle\frac{\rho_q}{2\omega}\rangle+
\sigma_{qg}\langle\frac{\rho_g}{2\omega}\rangle=
\frac{2\pi\alpha^2_s}{\mu^2}T^2,
\end{equation}
\begin{equation}
\sigma_{gq}\langle\frac{\rho_q}{2\omega}\rangle+
\sigma_{gg}\langle\frac{\rho_g}{2\omega}\rangle=
\frac{9}{4}\frac{2\pi\alpha^2_s}{\mu^2}T^2.
\end{equation}
The elastic energy loss of a fast parton inside a quark gluon
plasma at temperature $T$ is then given by,
\begin{equation}
\frac{dE_{el}}{dz}=C_2\frac{3\pi\alpha_s^2}{2\mu^2}T^2
\langle q^2_{\perp}\rangle. \label{eq:dedz2}
\end{equation}
For comparison, we plot this elastic energy loss in Fig.~\ref{fig4}.
In general, it is much smaller than the radiative energy loss
and has a weaker energy dependence (single logarithmic).

Using Eqs.~(\ref{eq:avq}), (\ref{eq:dedzf}) and (\ref{eq:xi}), the
total energy loss can be expressed as,
\begin{equation}
\frac{dE}{dz}=\frac{dE_{el}}{dz}+\frac{dE_{rad}}{dz}\approx
\frac{C_2\alpha_s}{\pi}\mu^2\ln\frac{3ET}{2\mu^2}\left(
\ln\frac{9E}{\pi^3T}+\frac{3\pi^2\alpha_s}{2\mu^2}T^2\right).
\label{eq:dedzt}
\end{equation}
It is interesting to note that both the elastic and radiative energy
loss have the same color coefficient $C_2$. For high energy partons,
the radiative energy loss is dominate over the elastic one. For
$E=30$ GeV, $T=300$ MeV, and  $\alpha_s\approx 0.3$,
the total energy loss for a propagating quark is $dE/dz\approx 3.6$ GeV/fm.
Only about 25\% of this amount comes from elastic energy loss.

\section{SUMMARY AND DISCUSSIONS}

We extended our previous derivation by
considering the  role of gauge invariance and target radiation in
the case of spin 1/2 quarks to improve
our estimate of radiative energy loss of a fast parton inside
a quark gluon plasma. Our main result [Eq. \ref{eq:dedz1})]
interpolates between the factorization
and Bethe-Heitler limits, and has unique nonabelian properties.
The factorization limit is of course consistent with the general
bound \cite{BRODS} imposed by the uncertainty principle,
but reveals peculiar energy and temperature dependence
of the mean square radiation transverse momentum
controlling that energy loss.
The total energy loss is very sensitive to the color screening scale
in the plasma. The
double logarithmic energy dependence of $dE_{rad}/dz$ is the result
of non-abelian aspects of the LPM effect in QCD.
The same effect should be responsible for
the limited gluon equilibration rate as discussed in Ref.~\cite{BMTW}.

Our derivation improves that in
\cite{GPTW,MGXW} in a number of ways. First, an
effective formation time $\tau_{\rm QCD}$ in QCD radiation was
used to  account for the color interference due to multiple
scatterings. The dependence of this effective formation time on the
color representation of the jet parton gave rise to the different
color factors, proportional to $C_2$, for the radiative
energy loss of a quark and gluon. In contrast both are
proportional to $C_A$ in the case of a single scattering.
Secondly, the gluon spectrum including  radiation from both
the jet line and the internal gluon line and
regulated consistently with  the requirement of gauge
invariance was used.  However, Eq. (\ref{eq:dedz1})
is still an idealization to the physically realizable
situation in nuclear collisions because
a number of strong assumptions were made in its derivation.
The strongest is the extrapolation of pQCD in a
regime $g\sim 1$ and the assumption
that the interaction range is small compared to the
color relaxation mean free path. We have therefore
left with an explicit dependence
on $\mu$ in $dE/dz$ since strong nonperturbative
variations of $\mu(T)$ occur in the vicinity of $T_c$.
The basic result that $dE/dz$ is proportional
to  $\mu^2$ is however very general and consistent with
the uncertainty bounds in\cite{BRODS}. Therefore, in a
separate paper \cite{GPW2}, we will investigate the phenomenological
consequences of interesting temperature
dependence of $\mu(T)$ suggested  by lattice calculations.

\acknowledgments
We would like to thank J.~Cleymans, J.~Knoll, and B.~M\"uller for
stimulating discussions.
This work was supported by the Director, Office of Energy
Research, Division of Nuclear Physics of the Office of High
Energy and Nuclear Physics of the U.S. Department of Energy
under Contract No. DE-AC03-76SF00098 and DE-FG02-93ER40764.

\appendix

\section*{}

The radiation amplitude within multiple scattering theory has been
derived in Ref. \cite{MGXW} in a general from. In this appendix we use
the radiation induced by double scattering as an example to demonstrate
how the general formula arises from the multiple scattering theory.
We consider the scattering of a high energy particle off potentials
as given by Eq.~(\ref{eq:v1}). For simplicity, let us first neglect the
color indices as in the case of QED. The amplitude for a single
scattering is then
\begin{equation}
  {\cal M}_{el}^{(1)}=2\pi i\delta(E_i-E_f)4\pi E f_1(E,t), \label{eq:ap1}
\end{equation}
\begin{equation}
f_1(E,t)=\frac{-i}{2\pi}\frac{g}{2Ei}\bar u_{\sigma_f}(p_f)
\not\!\!A({\bf q})u_{\sigma_i}(p_i)e^{-i{\bf q}\cdot{\bf x}},
\label{eq:ap2}
\end{equation}
where $E=E_i=E_f$ and the amplitude $f(E,t)$ is defined such that the
differential cross section  is given by
\begin{equation}
  \frac{d\sigma}{dt}=\pi |f(E,t)|^2, \label{eq:ap3}
\end{equation}
and $\sigma_i$, $\sigma_f$ are the initial and final polarizations which
should be averaged and summed over in the calculation of cross sections.
One can check that with the definition of $A({\bf q})$ in Eq.~(\ref{eq:v1}),
the above formula leads to $d\sigma/dt=4\pi\alpha_s^2/({\bf q}^2+\mu^2)^2$.

One can similarly write down the amplitude for double scatterings,
\begin{eqnarray}
{\cal M}_{el}^{(2)}=2\pi i\delta(E_i-E_f)(-g^2)&\int\frac{d^3\ell}{(2\pi)^3}
\bar u_{\sigma_f}(p_f)\not\!\!A({\bf p}_f-{\bf \ell})
\frac{\not\,\ell}{\ell^2+i\epsilon}\not\!\!A({\bf \ell}-{\bf p}_i)
 u_{\sigma_i}(p_i) \nonumber \\
&e^{-i({\bf \ell}-{\bf p}_i)
\cdot{\bf x}_1-i({\bf p}_f-{\bf \ell})\cdot{\bf x}_2},\label{eq:ap4}
\end{eqnarray}
where the energy conservation at each potential vertex sets the energy
of the internal line to $\ell^0=E=E_f$. Amplitudes involving backscattering
are suppressed at high energies because of the limited momentum transfer
that each potential can impart. When the mean free path (or distance
between two sequential scatterings) is much larger than the interaction
range ($\sim 1/\mu$) of the potential in Eq.~(\ref{eq:v1}), the singularity
in $A({\bf q})$ can be neglected and the integration over $\ell_z$ (with
respect to the $\hat{z}$ direction of
${\bf x}_{21}={\bf x}_2-{\bf x}_1=L\hat{z}+{\bf r}_{\perp}$) gives us
\begin{eqnarray}
{\cal M}_{el}^{(2)}&=&2\pi i\delta(E_i-E_f)\int\frac{d^2\ell_{\perp}}{(2\pi)^2}
\bar u_{\sigma_f}(p_f)(-g^2)\Gamma_{(2)} u_{\sigma_i}(p_i),\nonumber\\
\Gamma_{(2)}&=&\not\!\!A({\bf p}_f-{\bf p})
\frac{\not\!p}{2ip_z}\not\!\!A({\bf p}-{\bf p}_i)
e^{-i({\bf p}-{\bf p}_i)
\cdot{\bf x}_1-i({\bf p}_f-{\bf p})\cdot{\bf x}_2},\label{eq:ap5}
\end{eqnarray}
where $p=(E,\sqrt{E^2-\ell_{\perp}^2},{\bf \ell}_{\perp})$ is the
four-momentum of the internal line. One can derive the classical
Glauber multiple collision cross section from this amplitude by
averaging and summing over initial and final state ensemble of the
target\cite{MGXW}. In the limit of high energy and small angle scattering,
one can neglect the phase factor in the above equation and obtain
the amplitude [as defined in Eq.~(\ref{eq:ap1})],
\begin{equation}
f_2(E,t)=\frac{-i}{2\pi}\int d^2b\frac{1}{2!}[-\chi({\bf b},E)]^2
e^{i{\bf q}_{\perp}\cdot{\bf b}}, \label{eq:ap6}
\end{equation}
where $1/2!$ comes from the different ordering of the target potentials
and ${\bf q}_{\perp}={\bf p}_{f\perp}-{\bf p}_{i\perp}$ is the total
transverse momentum transfer due to the multiple scatterings. The
eikonal function $\chi_{\sigma_1,\sigma_2}({\bf b},E)$ is defined as the
Fourier transform of the single scattering amplitude (besides a
factor $i/2\pi$),
\begin{equation}
\chi_{\sigma_2,\sigma_1}({\bf b},E)\equiv -\int\frac{d^2q_{\perp}}{(2\pi)^2}
e^{-i{\bf q}_{\perp}\cdot{\bf b}}\frac{g}{2Ei}\bar u_{\sigma_2}(p_2)
\not\!\!A({\bf q}_{\perp}) u_{\sigma_1}(p_1). \label{eq:ap7}
\end{equation}
In the definition of the product of eikonal functions, summation over
the polarizations of the intermediate lines is implied,
\begin{equation}
\chi^n\equiv \sum_{\sigma_1,\cdots,\sigma_{n-1}}
\chi_{\sigma_f,\sigma_{n-1}}\chi_{\sigma_{n-1},\sigma_{n-2}}\cdots
\chi_{\sigma_1,\sigma_i}. \label{eq:ap8}
\end{equation}
One can generalize the double scattering amplitude to multiple
scatterings and sum them together to get the total amplitude,
\begin{eqnarray}
f(E,t)=\sum_{n=1}^{\infty}f_n(E,t)&=&\frac{-i}{2\pi}
\int d^2b \frac{1}{n!}[-\chi({\bf b},E)]^n e^{i{\bf q}_{\perp}\cdot{\bf b}}
\nonumber \\
&=&\frac{i}{2\pi}\int d^2b [1-e^{-\chi({\bf b},E)}]
e^{i{\bf q}_{\perp}\cdot{\bf b}}. \label{eq:ap9}
\end{eqnarray}
This is recognized as the elastic amplitude in the eikonal formalism. One
can generalize this to the case of Pomeron exchange so that one can obtain
both the total and inelastic cross sections for hadron-hadron
collisions\cite{WANG91,WANGTH}.

For radiation induced by double scatterings, there are contributions from
three different diagrams as illustrated in Fig.~\ref{fig2}. The total amplitude
can be written as,
\begin{equation}
{\cal M}_{rad}^{(2)}=2\pi i\delta(E_i-E_f-\omega)\int
\frac{d^2\ell_{\perp}}{(2\pi)^2}
\bar u(p_f)(-g^3)[\Gamma_a+\Gamma_b+\Gamma_c]u(p_i), \label{eq:ap10}
\end{equation}
\begin{eqnarray}
\Gamma_a=&\int\frac{d\ell_z}{2\pi}
\not\!\!A({\bf p}_f-{\bf\ell})\left[\frac{\not\,\ell}
{\ell^2+i\epsilon}\right]_{\ell^0=E}\not\!\!A({\bf\ell}+{\bf k}-{\bf p}_i)
\frac{\not{p}_i-\not{k}}{(p_i-k)^2}\not\!\varepsilon \nonumber\\
&e^{-i({\bf\ell}+{\bf k}-{\bf p}_i)\cdot{\bf x}_1
-i({\bf p}_f-{\bf\ell})\cdot{\bf x}_2}, \label{eq:ga}\\
\Gamma_b=&\int\frac{d\ell_z}{2\pi}
\not\!\!A({\bf p}_f+{\bf k}-{\bf\ell})\left[
\frac{\not\,\ell-\not{k}}{(\ell-k)^2+i\epsilon}\not\!\varepsilon
\frac{\not\,\ell}{\ell^2+i\epsilon}\right]_{\ell^0=E+\omega}
\not\!\!A({\bf\ell}-{\bf p}_i) \nonumber\\
&e^{-i({\bf\ell}-{\bf p}_i)\cdot{\bf x}_1
-i({\bf p}_f+{\bf k}-{\bf\ell})\cdot{\bf x}_2}, \label{eq:gb}\\
\Gamma_c=&\int\frac{d\ell_z}{2\pi}
\not\!\varepsilon\frac{\not{p_f}+\not{k}}{(p_i+k)^2}
\not\!\!A({\bf p}_f+{\bf k}-{\bf\ell})\left[\frac{\not\,\ell}
{\ell^2+i\epsilon}\right]_{\ell^0=E+\omega}
\not\!\!A({\bf\ell}-{\bf p}_i) \nonumber\\
&e^{-i({\bf\ell}-{\bf p}_i)\cdot{\bf x}_1
-i({\bf p}_f+{\bf k}-{\bf\ell})\cdot{\bf x}_2}, \label{eq:gc}
\end{eqnarray}
where we have denoted $E=E_f=E_i-\omega$. Similarly as in double elastic
scattering, one can integrate over $\ell_z$ and obtain,
\begin{eqnarray}
\Gamma_a=&\not\!\!A({\bf p}_f-{\bf p})
\frac{\not{p}}{2ip_z}\not\!\!A({\bf p}+{\bf k}-{\bf p}_i)
\left(-\frac{\varepsilon\cdot p_i}{k\cdot p_i}\right)
e^{-i({\bf p}+{\bf k}-{\bf p}_i)\cdot{\bf x}_1
-i({\bf p}_f-{\bf p})\cdot{\bf x}_2}, \label{eq:ga1} \\
\Gamma_c=&
\not\!\!A({\bf p}_f+{\bf k}-{\bf p}^{\prime})
\frac{\not{p}^{\prime}}{2ip_z^{\prime}}
\not\!\!A({\bf p}^{\prime}-{\bf p}_i)
\left(\frac{\varepsilon\cdot p_f}{k\cdot p_f}\right)
e^{-i({\bf p}^{\prime}-{\bf p}_i)\cdot{\bf x}_1
-i({\bf p}_f+{\bf k}-{\bf p}^{\prime})\cdot{\bf x}_2}, \label{eq:gc1}
\end{eqnarray}
where we have used Dirac equations for the spinors and taken the
soft radiation limit (neglecting terms like $\not\!\varepsilon\not\!{k}$).
Notice that the four momenta for the intermediate line between two
scatterings before and after the radiation are different,
\begin{eqnarray}
p&=&(E,\sqrt{E^2-\ell_{\perp}^2},{\bf\ell}_{\perp}),\nonumber\\
p^{\prime}&=&(E+\omega,\sqrt{(E+\omega)^2-\ell_{\perp}^2},
{\bf\ell}_{\perp}).
\end{eqnarray}
Define $\phi_D=-({\bf p}-{\bf p}_i)\cdot{\bf x}_1
-i({\bf p}_f-{\bf p})\cdot{\bf x}_2=p_zL+
{\bf\ell}_{\perp}\cdot{\bf r}_{\perp}+{\bf p}_i\cdot{\bf x}_1
-{\bf p}_f\cdot{\bf x}_2$ as the phase factor for double elastic
scattering, the corresponding phase factors in the above two amplitudes
for induced radiation become,
\begin{equation}
\phi_a=\phi_D+ik\cdot x_1,\;\;\; \phi_c=\phi_D+ik\cdot x_2
\end{equation}
where we used $p^{\prime}_z\simeq p_z+\omega/v_z$, $v_z=p_z/E$
and we defined the time components of the four-coordinates as
$t_1=0$, $t_2=L/v_z$. These two terms can be identified with the initial
radiation of the first scattering and the final state radiation
of the second scattering, respectively.

In the amplitude of the radiation from the middle line, $\Gamma_b$, we can
see that there are singularities in both of the two propagators. They should
both contribute to the integration over $\ell_z$. To complete the
integration, we may use the identity (with $k^2=0$),
\begin{equation}
[(\ell-k)^2+i\epsilon]^{-1}[\ell^2+i\epsilon]^{-1}
=\frac{[(\ell-k)^2+i\epsilon]^{-1}}{2k\cdot(\ell-k)}
-\frac{[\ell^2+i\epsilon]^{-1}}{2\ell\cdot k}.
\end{equation}
With $\ell^0=E+\omega$, the singularity in the second term put the
internal line before the radiation vertex on mass-shell with
four-momentum $p^{\prime}$. This contribution  corresponds to
the initial state radiation of the second scattering. For the first
term in the above equation, we can make a variable change,
$\ell^{\prime}=\ell-k$. With  $\ell^{\prime}_0=E$, the internal line
after the radiation vertex is put on shell with four-momentum $p$.
This term can be identified as the final state radiation of the
first scattering. Using identities,
\begin{eqnarray}
\not{p}\not\!\varepsilon(\not{p}+\not{k})&=&2\varepsilon\cdot p \not{p},\\
(\not{p}^{\prime}-\not{k})\not\!\varepsilon\not{p}^{\prime}&=&
2\varepsilon\cdot p^{\prime}\not{p}^{\prime},
\end{eqnarray}
one has the amplitude of the radiation
from the middle line between two scatterings,
\begin{eqnarray}
\Gamma_b&=&\not\!\!A({\bf p}_f-{\bf p})
\frac{\not{p}}{2ip_z}\not\!\!A({\bf p}+{\bf k}-{\bf p}_i)
\left(\frac{\varepsilon\cdot p}{k\cdot p}\right)
e^{i\phi_D+ik\cdot x_1} \nonumber \\
&+&\not\!\!A({\bf p}_f+{\bf k}-{\bf p}^{\prime})
\frac{\not{p}^{\prime}}{2ip_z^{\prime}}
\not\!\!A({\bf p}^{\prime}-{\bf p}_i)
\left(-\frac{\varepsilon\cdot p^{\prime}}{k\cdot p^{\prime}}\right)
e^{i\phi_D+ik\cdot x_2}, \label{eq:gb1}
\end{eqnarray}
Summing all contributions together, we have the total amplitude for
radiation induced by double scatterings,
\begin{eqnarray}
\Gamma_a+\Gamma_b+\Gamma_c&=&-\not\!\!A({\bf p}_f-{\bf p})
\frac{\not{p}}{2ip_z}\not\!\!A({\bf p}+{\bf k}-{\bf p}_i)
\left(\frac{\varepsilon\cdot p_i}{k\cdot p_i}-
\frac{\varepsilon\cdot p}{k\cdot p}\right)e^{i\phi_D+ik\cdot x_1}
\nonumber \\
&-&\not\!\!A({\bf p}_f+{\bf k}-{\bf p}^{\prime})
\frac{\not{p}^{\prime}}{2ip_z^{\prime}}
\not\!\!A({\bf p}^{\prime}-{\bf p}_i)
\left(\frac{\varepsilon\cdot p^{\prime}}{k\cdot p^{\prime}}-
\frac{\varepsilon\cdot p_f}{k\cdot p_f}\right)
e^{i\phi_D+ik\cdot x_2}
\end{eqnarray}
which has two distinct contributions induced by each scatterings. In
the high energy and small angle scattering limit, we can neglect
the small momentum transfer at each scattering as compared to the
beam energy $E$. We also assume the soft radiation limit in which
$k_{\perp}\ll q_{\perp}$, with $q_{\perp}$ being the transverse momentum
transfer at each elastic scattering. Substituting the above expression
back into Eq.~(\ref{eq:ap10}), the total amplitude factorizes as
\begin{equation}
{\cal M}_{rad}^{(2)}\approx {\cal M}_{el}^{(2)}i{\cal R}_2^{\rm QED},
\end{equation}
with the double elastic amplitude ${\cal M}_{el}^{(2)}$ given by
Eq.~(\ref{eq:ap5}) and the radiation amplitude ${\cal R}_2^{\rm QED}$
by Eq.~(\ref{eq:radQED}).

In the case of gluon radiation in QCD, we will have different color
matrices for different diagrams in Fig.~\ref{fig2}. With the color indices
as defined in Fig.~\ref{fig2}, the corresponding color factors for $\Gamma_a$,
$\Gamma_b$, and $\Gamma_c$ are, respectively,
\begin{eqnarray}
\left(T^{a_2}T^{a_1}T^b\right)_{BB^{\prime}}T^{a_1}_{A_1A_1^{\prime}}
T^{a_2}_{A_2A_2^{\prime}}, \nonumber \\
\left(T^{a_2}T^bT^{a_1}\right)_{BB^{\prime}}T^{a_1}_{A_1A_1^{\prime}}
T^{a_2}_{A_2A_2^{\prime}}, \nonumber \\
\left(T^bT^{a_2}T^{a_1}\right)_{BB^{\prime}}T^{a_1}_{A_1A_1^{\prime}}
T^{a_2}_{A_2A_2^{\prime}}.
\end{eqnarray}
When taking the high energy limit, all the contributions have a common
momentum dependence, $\epsilon\cdot p_i/k\cdot p_i\simeq
\epsilon\cdot p/k\cdot p \simeq \epsilon\cdot p_f/k\cdot p_f
\simeq 2\vec{\epsilon}_{\perp} \cdot{\bf k}_{\perp}/k^2_{\perp}$.
One can immediately arrives at Eq.~(\ref{eq:radQCD}).

\noindent
\begin{figure}
\caption{Diagrams for induced gluon radiation from a single $qq$ scattering.}
\label{fig1}
\end{figure}

\begin{figure}
\caption{Diagrams for gluon radiation from the quark line induced by
double scatterings.}
\label{fig2}
\end{figure}

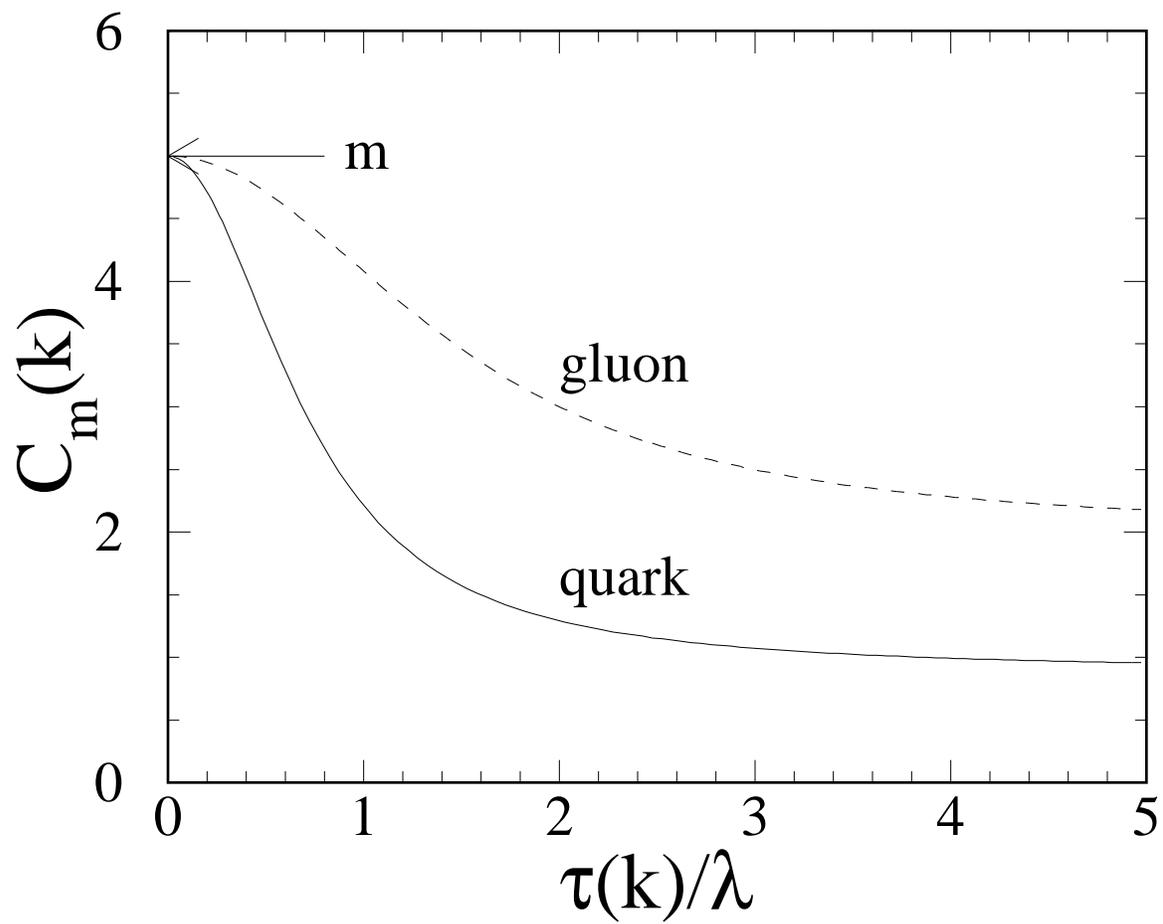
\begin{figure}
\caption{The radiation formation factor $C_m(k)$, $m=5$, as a function of
        $\tau(k)/\lambda=2\cosh(y)/k_{\perp}\lambda$ for quarks (solid) and
        gluons (dashed).}
\label{fig3}
\end{figure}

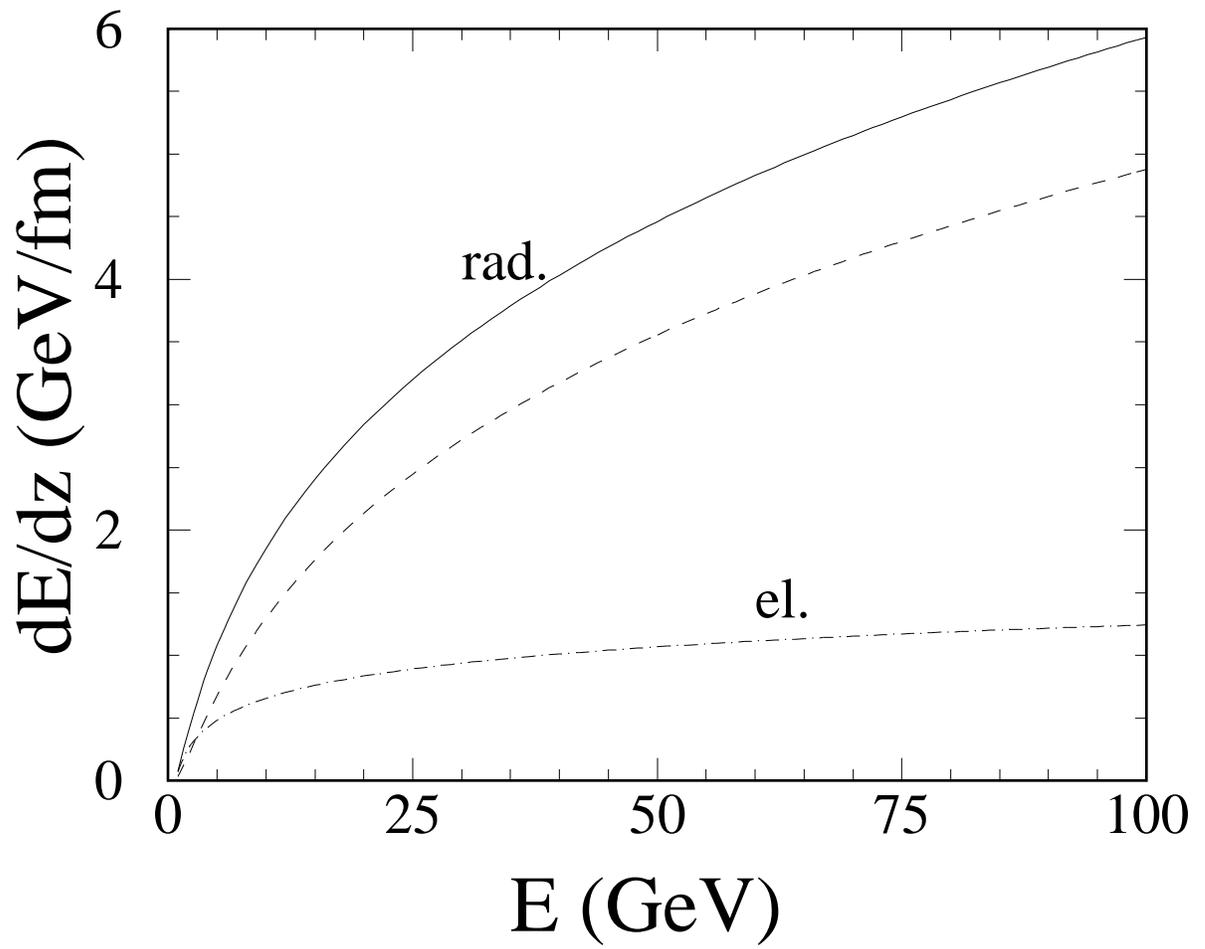
\begin{figure}
\caption{The energy dependence of energy loss, $dE/dz$, of a quark with
        energy $E$ inside a quark-gluon plasma at temperature $T=300$ MeV.
        A weak coupling $\alpha_s=0.3$ is used. The solid line is the full
        expression and the dashed line is the factorization limit of the
        radiative energy loss. The dot-dashed line is the elastic energy
        loss.}
\label{fig4}
\end{figure}

\end{document}